\documentclass[AMA,STIX2COL, preprint]{WileyNJD-v2}
\usepackage{moreverb}
\usepackage{textgreek}
\newcommand\BibTeX{{\rmfamily B\kern-.05em \textsc{i\kern-.025em b}\kern-.08em
T\kern-.1667em\lower.7ex\hbox{E}\kern-.125emX}}
\newcommand{\new}[1]{\textcolor{black}{#1}}
\newcommand{\rev}[1]{\textcolor{black}{#1}}

\articletype{}%

\received{}
\revised{}
\accepted{}

\begin{document}

\title{Magnetic Tilting in Nematic Liquid Crystals driven by Self-Assembly}

\author[1,2]{Martin H\"ahsler}

\author[3]{Hajnalka N\'adasi}

\author[3]{Martin Feneberg}

\author[4]{Sebastian Marino}

\author[4]{Frank Giesselmann}

\author[1,2]{Silke Behrens*}

\author[3]{Alexey Eremin*}

\authormark{H\"ahsler \textsc{et al}}

\address[1]{Institute of Catalysis Research and Technology,
Karlsruhe Institute of Technology,
76344 Eggenstein-Leopoldshafen, Germany}

\address[2]{Institute of Inorganic Chemistry, Ruprecht-Karls University,
69120 Heidelberg, Germany}

\address[3]{Institute of Physics, Otto von Guericke University Magdeburg,
39106 Magdeburg, Germany}

\address[4]{Institute of Physical Chemistry,
University of Stuttgart
70569 Stuttgart, Germany}

\corres{*Alexey Eremin, Otto von Guericke University, Inst. of Physics, 39016 Magdeburg, Germany \email{alexey.eremin(at)ovgu.de}, \\
*Silke Behrens, Institute of Catalysis Research and Technology
Karlsruhe Institute of Technology
76344 Eggenstein-Leopoldshafen, Germany \email{silke.behrens(at)kit.edu}}

\abstract[Abstract]{Self-assembly is one of the crucial mechanisms allowing to design multifunctional materials. Soft hybrid materials contain components of different nature and exhibit competitive interactions which drive self-organisation into structures of a particular function. Here we demonstrate a novel type of a magnetic hybrid material where the molecular tilt can be manipulated through a delicate balance between the topologically-assisted colloidal self-assembly of \rev{magnetic nanoparticles} and the anisotropic molecular interactions in a liquid crystal matrix.
}

\keywords{hybrid materials, liquid crystals, magneto-optical materials, magnetic materials, self-assembly }

\maketitle

\footnotetext{\textbf{Abbreviations:} none}

% \section{Introduction}\label{sec1}
Soft materials with stimuli-responsive behavior generated significant interest over the past few years for constructing smart and functional systems.\cite{Liu:2019co, Zhang:2019id} In particular, hybridization of liquid crystals (LC) with magnetic nanoparticles (MNPs) opens new avenues for designing novel magneto-responsive materials and devices for transparent magnets,\cite{Mertelj:2014kv} actuators,\cite{Wang:2014em} sensors,\cite{Wang:2016iz} data storage,\cite{Hu:2010ip} and direct visualization of magnetic fields.\cite{Rupnik:2015bi} LCs are characterized by the combination of fluidity of ordinary liquids with the anisotropic electro- and magneto-optical properties of crystalline materials.\cite{deGennes:1992jj} The symmetry of nematic LCs (N-LC) is uniaxial where rod-shaped molecules (mesogens) orient along a common direction, the director \textbf{n}. Optical properties of LCs (e.g. in liquid crystal displays) are typically controlled by the (re)orientation of mesogens in electric fields. As an alternative to electrical control, N-LCs can also be manipulated by magnetic fields; however, due to the low diamagnetic anisotropy of the mesogens, magnetic actuation requires strong magnetic fields. The contactless nature of magnetic manipulation and versatility of magnetic, magneto-optical and magneto-electro-optical effects are extremely attractive and have motivated numerous studies on magnetic LC hybrid materials in this field. Doping of N-LCs with MNPs, for example, not only led to enhanced sensitivity to the magnetic field in ferronematics\cite{Brochard:1970kp,Appel:2017hs} but also to ferromagnetic nematics with residual magnetization, which were only recently discovered in dispersions of magnetic platelets.\cite{Mertelj:2014kv,Rupnik:2015bi,{Mertelj:2018gq}} The highly interesting, stimulus-responsive properties of these materials are based on a complex interplay between magnetic dipolar and LC-mediated elastic interactions and coupling of the nematic director \textbf{n} to the magnetization \textbf{M} of the MNPs. These effects not only critically influence the stability of the material but also its magnetic, magneto- and electro-optical properties and response dynamics, but are yet not well understood.\cite{Brochard:1970kp} While typical ferronematics with weak anchoring of the nematic director at the MNP surface behave as superparamagnetic even though the MNPs are ferromagnetic, in so-called ferromagnetic nematics the director strongly anchors at the MNPs and stabilizes the ferromagnetic colloidal order.

The interface between MNP and LC host (i.e. the chemical surface properties of the MNPs) are of utmost importance. Dendritic ligands with (pro)mesogenic entities not only kinetically stabilize the MNPs but also reduce both the disturbance of the LC order in the proximity of the MNPs as well as the tactoidal distortion and equatorial depletion of the the flexible ligand shell, preventing MNP agglomeration and phase separation.\cite{Koch:2020bt,Draper:2011ej} Additionally, the (pro)mesogenic ligands critically control the coupling strength between the LC host and the magnetic moments \textbf{m} of the MNPs. Recently, also a remarkable, through-space amplification of chirality was observed for minute amounts of nanorods capped with chiral ligands in an achiral N-LC.\cite{Nemati:2018cp} Self-assembly of embedded nanoparticles into 3D structures may further affect the global ordering of the LC. The formation of modulated phases often occurs as a result of the competition between two or more ordering mechanisms favoring different equilibrium states. Various systems where composition,\cite{Seul_Science_1995} molecular tilt, and surface curvature \cite{Chen_PRE_1995, Kamien:2001uo} are modulated are found in nature. However, hybrid materials of a nematic LC-phase and dendronized MNPs, where the tilt of the nematic director can be adjusted by magnetically induced self-assembly and with controlable magnetic properties remain unknown.

%However, whether it is possible to obtain a hybrid material of a nematic LC-phase and dendronized MNPs where the tilt of the nematic director can be adjusted by magnetic induced self-assembly and with control of the magnetic properties exclusively remains unknown.

For the first time, we report on a magnetic LC material where self-assembly of MNP modulates the tilt of the director \textbf{n}. Consequently, some of the important issues we are addressing here are the coupling of the magnetic MNP subphase to the LC matrix and the magneto- and electro-optical behavior of the N-LC hybrid materials. Stable dispersions of MNPs in a N-LC are achieved by functionalizing the MNPs with a (pro)mesogenic dendron. In the magnetic field, the mesogens align tilted to the direction of the magnetic field. 
This effect can not be explained by the common model of the magnetic Fr\'eedericksz transition for ferronematics with uniform MNP distribution in the N-LC matrix and indicates the self-assembly of functinalized MNPs into superstructures, which in turn seem to induce the inclined orientation of the nematic director \textbf{n}. Detailed characterization of the magneto-/electro-optical properties and structure by impedance spectroscopy, polarized optical microscopy (POM), small- and wide-angle X-ray scattering (SAXS, WAXS), and Raman spectroscopy is reported. 

\begin{figure}[hbt]
\centering
	\includegraphics[width=0.95\columnwidth]{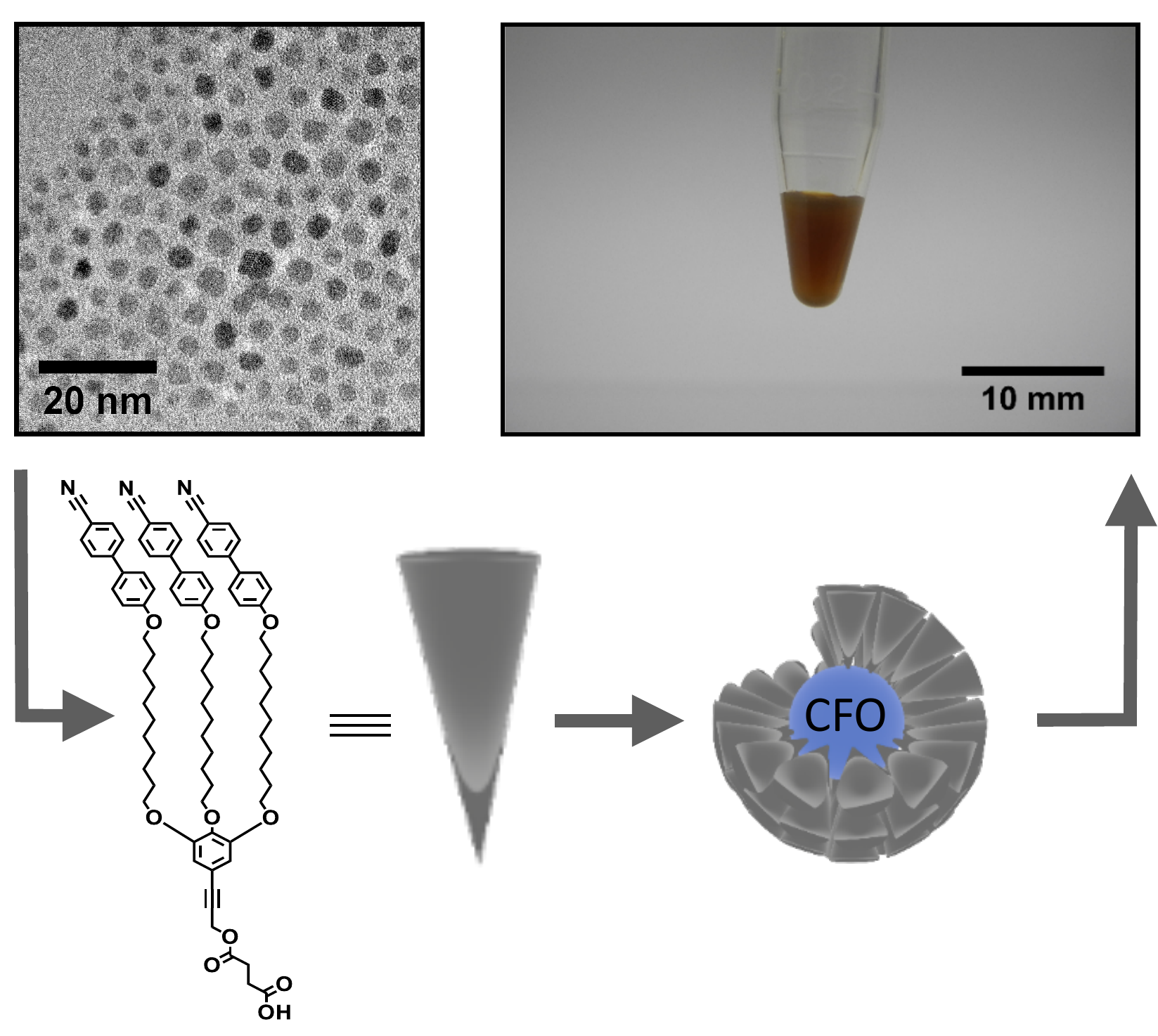}
	\caption{(top left) TEM images of the MNPs; (bottom)  schematics of a functionalized nanoparticle; (top right) a dispersion of the MNPs in 5CB (Dend-MNP-5CB-hybrid material). 
		\label{fig:fig1} 
	}
\end{figure}

%\section{Results}
%\subsection{Materials}
In a first step, the synthesis of MNPs and their chemical surface properties were addressed to prepare a magnetic LC hybrid material with high colloidal stability in the N-LC host and strong sensitivity to the magnetic field. 4-cyano-4'-pentylbiphenyl (5CB), a commercial N-LC \rev{with the calorimetry data shown in Fig.~S16}, was employed as N-LC host. Uniform, 4.6 nm CFO-MNPs were synthesized by reaction of Fe(acetylacetonate)\textsubscript{3} and Co(acetylacetonate)\textsubscript{2} with 1,2-hexadecanediol in the presence of oleic acid and oleylamine (see Supporting Information).\cite{Hahsler:2020ci,Sun:2004jm} Cobalt ferrite (CFO) reveals a high chemical and physical stability together with an unusually high magnetic anisotropy ($\sim 2 \times 10^5$ J/m\textsuperscript{3}) and a superparamagnetic limit at $\sim$8 nm. As previously shown by us, aliphatic, (pro)mesogenic ligands disperse very small MNPs in 5CB, but colloidal stability decreases with increasing MNP size and is very limited for MNPs larger than approximately 3 nm.\cite{Appel:2017hs} In this work, a sterically demanding, (pro)mesogenic ligand with dendritic structure (see \new{Figures S1, S2}, Supporting Information) was introduced to functionalize the surface of the 4.6 nm size MNPs and ensure their colloidal stability in the N-LC host. The nature of the ligand critically influences the coupling to the surrounding LC host. A ligand exchange reaction of 4.6 nm CFO-MNPs with a dendron yielded dendronized MNPs (Dend-MNPs).\cite{Hahsler:2019kt} Dend-MNPs were added to the 5CB host, sonicated at 60 $^{\circ}$C and quenched to the N-LC phase
(\textit{T} = 25 $^{\circ}$C). After removal of larger aggregates \textit{via} centrifugation, the homogeneous Dend-MNP-5CB-hybrid material was obtained (Figure~\ref{fig:fig1}). No residual (micro)aggregates were detected by optical microscopy. The size  of the MNPs and the clearing point of the N-LC were not affected by the
dispersion step ($\Delta d=0.04$~nm; $\Delta T=0.0^{\circ}$C). 

\begin{figure}[hbt]
\centering
	\includegraphics[width=0.8\columnwidth]{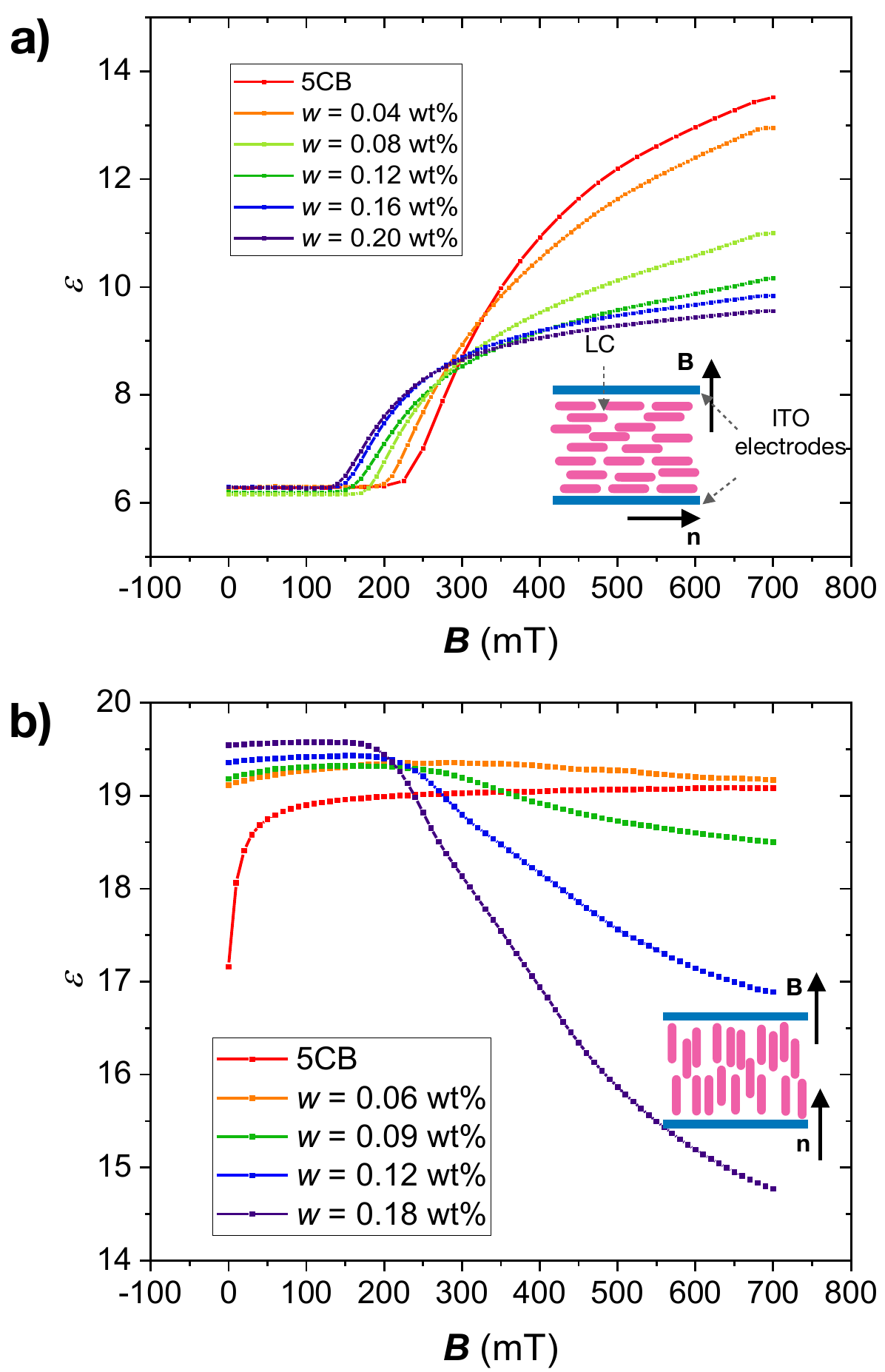}
	\caption{Electro and magneto-optical behavior of the ferronematic Dend-MNP-5CB-hybrid material: a) magnetic FT in a planar 25 \textmu m cell; b) Dielectric permittivity measured in a homeotropic 10 \textmu m cell as a function of the applied magnetic field. The sketches show the initial device geometry and the orientation of the applied fields. 
		\label{fig:fig2} 
	}
\end{figure}

%\subsection{Results}
The coupling effect of the magnetic subphase with the LC matrix strongly affects the magneto-optical behavior of a hybrid LC material. The Fr\'eedericksz transition (FT) constitutes the response of the nematic director to an applied electric or magnetic field, where the dielectric and diamagnetic torques counteract the elastic torque of the nematic director. \cite{deGennes:1995vg, {Brochard:1970kp}} Our ferronematic (Dend-MNP-5CB-hybrid material) and the non-doped LC show nearly identical behavior of the critical voltage $U_c$ (\new{Figure S5}, Supporting Information). The negligible difference in electro-optical behavior indicates that the Dend-MNPs in the nematic host 5CB do not significantly affect neither the order parameter $S$  nor the splay elastic constant by an $E$-field. By contrast, the realignment of the nematic director \textbf{n} in a magnetic field becomes very sensitive to the presence of the MNPs (Figure ~\ref{fig:fig2} and Figure \new{S6}, Supporting Information). A magnetic field applied perpendicular to the cell plane destabilizes the planar state even further by shifting the magnetic FT threshold (Figure~\ref{fig:fig2}a). The critical field $B_c$ for the magnetic FT decreases with the increasing mass fraction of MNPs. Thus, the ferronematic exhibits a \emph{positive} director-magnetization coupling favoring parallel alignment of \textbf{n} and \textbf{M}.

Although such a destabilization effect, attributed to the surface anchoring of the director at the MNP surface, has also been observed in other ferronematic systems,\cite{Kopcansky:2005cm, Tomasovicova:2008iq, Kopcansky:2008dm} the character of the transition in this system is very different. For an increasing magnetic field, we observe a systematic decrease \rev{of the} dielectric permittivity tensor component $\varepsilon_{||}$ parallel to the director.

Experiments combining \textbf{E}- and \textbf{B}-fields shed more light on the peculiar behavior of the ferronematic. \new{Figure~S6 a}, Supporting Information shows the electric FT for $w = 0.20$ wt$\%$ in a magnetic field bias aligned parallel to the unperturbed alignment direction of the (\textbf{B} $||$ \textbf{n} ${\perp}$ \textbf{E}). 
The threshold $U_c$ shifts to the lower voltage contradicting the conclusion of the \emph{positive} \textbf{n}-\textbf{M} coupling from the observation of the pure magnetic FT. 
Here, the magnetic field, parallel to the director \textbf{n}, has a destabilizing tendency. Also, the shape of the curve  $\varepsilon_{\mathrm{eff}}(U)$ deviates from the shape determined by the effective electromagnetic torque. 
This unusual step-like behaviour is stronger pronounced at higher mass fractions of MNPs (Figure S1, Supporting Information). When the magnetic field is applied parallel to the electric field in a planarly aligned nematic, the threshold of the electric FT decreases and the transition softens (Figure~15b, Supporting Information). This suggests that the magnetic field has a destabilizing effect even in this geometry.

Since the magnetic and the electric torques scale with critical fields as $B^2$ and $U^2$, respectively, it is useful to plot the threshold values in a $U^2/B^2$ diagram for \textbf{B} {\textbar}{\textbar} \textbf{n} ${\perp}$ \textbf{E} configuration (\new{Figure~S6 c}, Supporting Information). For the nematic host, the experimental points collapse on a straight line with a slope determined by the ratio of the dielectric to diamagnetic anisotropies of the mesogens.\cite{Appel:2017hs} In the case of the hybrid material, \textbf{B} has a destabilizing effect despite the positive diamagnetic anisotropy. \textit{U}\textit{\textsuperscript{2}}\textit{(B}\textit{\textsuperscript{2}}\textit{)} is nonlinear and exhibits a saturating behavior at a high \textbf{B}. In the case of the electric FT in a magnetic bias applied parallel to the electric field (\textbf{n} ${\perp}$\textbf{ B} {\textbar}{\textbar} \textbf{E}), the same destabilizing tendency was observed (\new{Figure~S6} b, Supporting Information). The transition sets off at lower voltages and  $\varepsilon _{||}$ decreases with increasing magnetic field. To test this assumption that the dielectric permittivity decreases along the magnetic field, we considered a different geometry, where the initial state is homeotropic and both fields are applied parallel to the director \textbf{n} {\textbar}{\textbar}\textbf{ B} {\textbar}{\textbar} \textbf{E }(Figure~\ref{fig:fig2}b). No change of the director orientation is expected in such a configuration. Indeed, the effective dielectric permittivity  $\varepsilon _{\mathrm{eff}}=\varepsilon _{||}$  of the undoped nematic displays no significant change. In the absence of a magnetic field, the same is true for the Dend-MNP-5CB-hybrid material.  In a magnetic field, however, we observe a substantial reduction of  $\varepsilon _{||}$ in agreement with the measurements in the planar cell (Figure~\ref{fig:fig2}a). This effect strongly depends on the concentration of the MNPs as shown in Figure S5, Supporting Information. The decrease of the  $\varepsilon _{||}$  can be related to a tilt of the mesogens or a decrease of the orientational order. The tilt of the molecules would affect the alignment of the mesogens with respect to the director \textbf{n} of the undoped nematic. The decrease of the orientational order would be attributed to a global distortion of the nematic order at the nanoscale. To explain the peculiar behavior of  $\varepsilon _{||}$, we investigated the orientational order parameter and molecular tilt using X-ray scattering and Raman spectroscopy.

%\subsubsection{Structure}

\begin{figure}[hbt]
\centering
	\includegraphics[width=0.9\columnwidth]{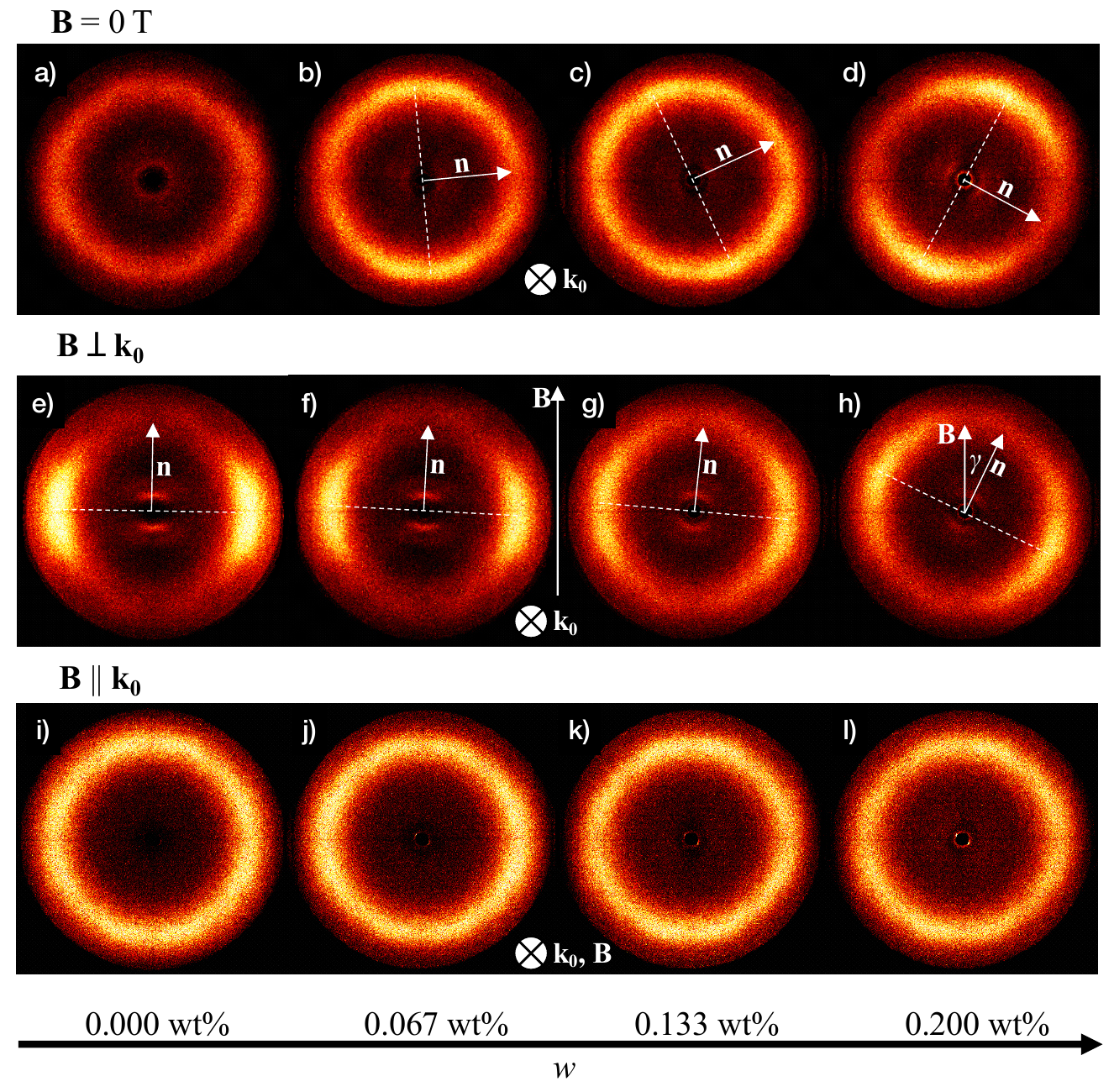}
	\caption{2D wide-angle X-ray scattering patterns of 5CB with MNP concentration of 0.000 wt$\%$, 0.067 wt$\%$, 0.133 wt$\%$ and 0.200 wt$\%$ in the nematic phase at $T = 25^{\circ}$C. Each row illustrates 2D-XRD-patterns of a certain setup: without an applied \textbf{B}-field in a-d, a permanent applied magnetic field orthogonal to the incoming X-ray beam $\mathbf{k_0}$ in e-h and the samples pre-oriented in an magnetic field of 800 mT and measured with an orientation $\mathbf{k_0}$ parallel to \textbf{B} in i- l.
		\label{fig:fig3} 
	}
\end{figure}

%Raman depolarisation ratios in the ferronematic material are nearly the same as in the pure 5CB (see Supporting Information). This suggests that the magnetic field does not affect the orientational order parameter. 
Small- (SAXS) and wide-angle (WAXS) X-ray scattering gives  insight into the orientational order and correlations of the mesogens in the nematic phase. In the absence of an external field, the scattering diagram shows no orientation for the pure LC. Surprisingly, the ferronematic samples exhibit pre-oriented patterns even without an external field (Figure~\ref{fig:fig3}a - d and \new{S7, Supporting Information}). The director orientation found by WAXS is arbitrarily chosen and appears to be independent on the capillary (substrate) orientation. Application of a magnetic field results in the molecular alignment. The WAXS scattering maxima remain on the equator of the pattern in the pure LC, suggesting that \textbf{n} {\textbar}{\textbar} \textbf{B} (Figure~\ref{fig:fig3}e). In the ferronematic, on the other hand, the director becomes tilted away from the direction of \textbf{B} (Figure~\ref{fig:fig3}f - h). The tilt direction is chosen arbitrary and depends on the position in the sample. The tilt angle $\gamma$, however, depends on the mass fraction of MNPs (\new{Figure S8, Supporting Information}). The scattering patterns with X-Ray beam $\mathbf{k_0}\parallel \mathbf{B}$ show an isotropic intensity distribution indicating no preferred in-plane order of the tilt (Figure~\ref{fig:fig3}i - l).
Considering the scattering profiles $I(\chi)$ with the orientation $\mathbf{k_0} \perp \mathbf{B}$ in (\new{Figure S8, Supporting Information}), the director tilt $\gamma $ to the magnetic field \textbf{B} increases from 1.2$^\circ$ to 25.5$^\circ$ with increasing MNP mass fraction. In addition, \new{Figure S8, Supporting Information} reveals peak broadening and a reduction of the scattering intensity with increasing MNP concentration. This broadening indicates a reduction of the effective nematic order in \textbf{B}. 
In the isotropic phase, the intensity distribution in the 2D-XRD-patterns remains uniform, suggesting no influence of the MNPs on the molecular alignment above the clearing point (\new{Figure S10, Supporting Information}). The scattering from the MNPs appears at small angles (in SAXS). The X-ray diagrams show a strong increase of the forward scattering intensity caused by the MNP form factor in ferronematic samples (\new{Figures S11 and S12, Supporting Information}). Furthermore, no scattering was observed in the SAXS-region of pure 5CB (\new{Figure S12}, Supporting Information). The azimuthal dependence of the scattering profile shows an anisotropy suggesting that the particles form aligned structures. Furthermore, no scattering was observed in the SAXS-region of pure 5CB.
%An additional maximum located at $q{\approx}$ 0.5 nm\textsuperscript{{}-1 }corresponds to a periodicity of 13 nm or the double of the NP diameter (7.5 nm) (\new{Figures S11 and S12, Supporting Information}). Furthermore, no scattering was observed in the SAXS-region of pure 5CB (\new{Figure S12}, Supporting Information). The azimuthal dependence of the scattering profile shows an anisotropy suggesting that the particles form aligned structures. Furthermore, no scattering was observed in the SAXS-region of pure 5CB.

% (cf. Figure S2, Supporting Information). 
  Thus, it can be concluded that the origin of this scattering is the appearance of the MNPs with a size of about 7.5 nm. It is also of no importance whether the nanoparticles are embedded in a nematic or even an isotropic environment as illustrated in \new{Figure S12 in the Supporting Information}, the differences in the scattering profiles $I(q)$ are negligible.

%\subsection{Discussion}
As demonstrated above, the behavior in a magnetic field does not have a straightforward explanation by Burylov-Raikher theory of ferronematics, \cite{Burylov:1995gc,Burylov:1995ie,Burylov:1990jz} where a magnetic coupling to ferromagnetic particles is assumed. The reduction of the effective dielectric permittivity occurs in the direction of the magnetic field independently of how the electric field was applied. A reasonable explanation could be a reduction of the orientational order parameter \textit{S} caused by the presence of the MNPs. However, the birefringence, which is proportional to \textit{S}, does not change significantly. Frank elastic constants in the ferronematic remain nearly the same as in the pure LC host. Also, the Raman scattering data do not show any markable effect of the MNPs on the orientational order parameter (\new{Figure S13, Supporting Information}). Instead, the director forms a macroscopic tilt to the magnetic field direction, as demonstrated by X-ray scattering. Molecular tilt explains the behavior of the effective dielectric permittivity in cells with planar and homeotropic conditions in a magnetic field. The tilted configuration appears to be stable in the limit of high \textbf{B}. 

\begin{figure}[hbt]
\centering
	\includegraphics[width=0.9\columnwidth]{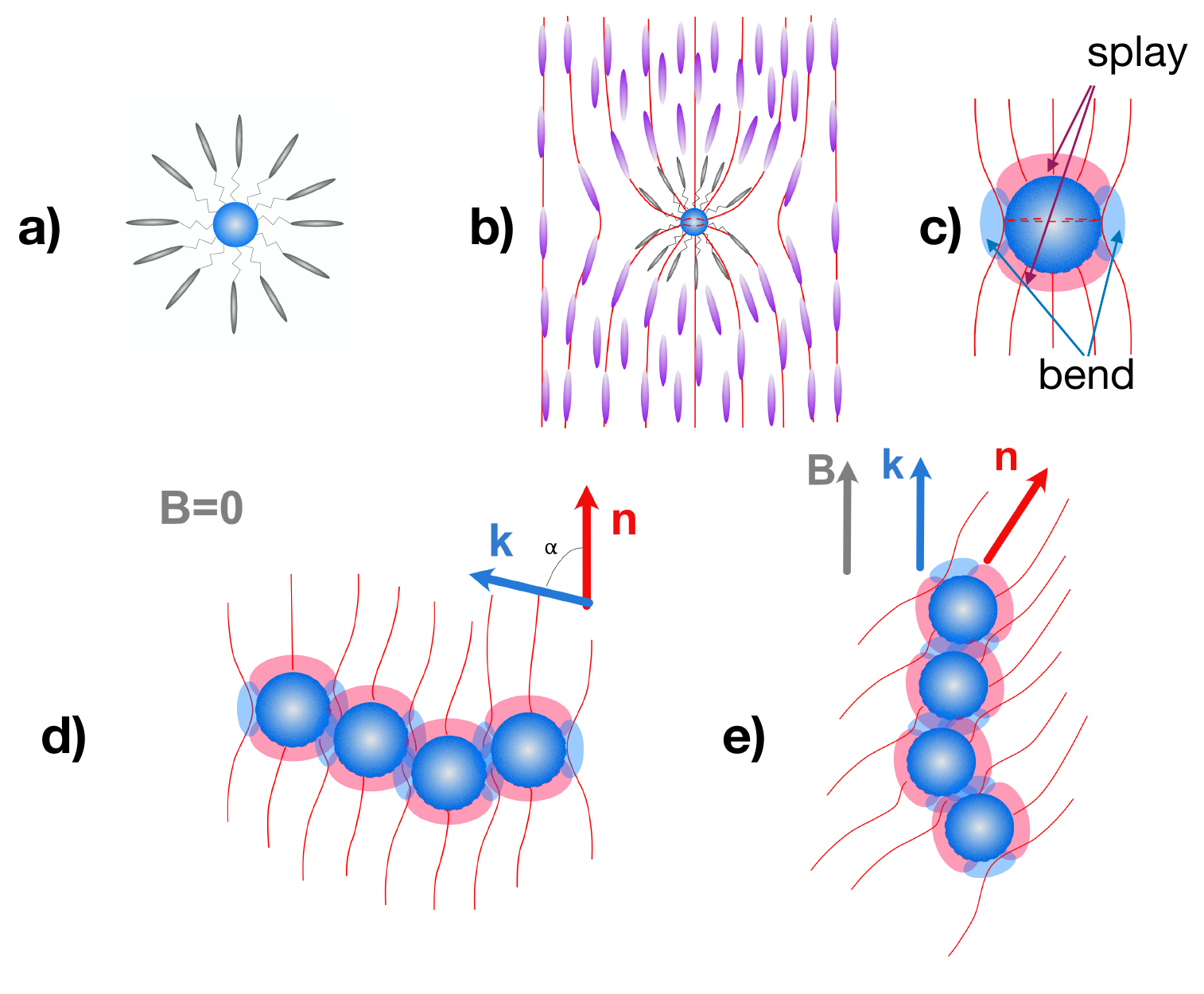}
	\caption{\rev{Schematics illustrating the self-assembly mechanism of the MNPs and tilt stabilization in the case of infinitely strong anchoring: a) the dendron branches are radially aligned at the particle core; in the LC, the MNPs form a quadrupole with a splay deformations at the poles and the bend deformation at the equator b) and c). Director-mediated interactions result in the formation of chains in a uniform director field d). The chains align along the field resulting in an oblique alignment of the director e).}
		\label{fig:fig4} 
	}
\end{figure}
Such an equilibrium requires an additional stabilization mechanism preventing the reorientation of the mesogens via the anisotropic diamagnetic coupling. The characteristic length-scale should lie in a mesoscale range but remain long enough not to disturb the microscopic nematic order. To explain the tilt, we propose a model where the MNPs self-assemble into chainlike structures which exhibit a sufficiently strong oblique anchoring to the nematic director. The key element of the model is the stabilization via topological constraints on the director field. Indeed, the magnetic interactions between the superparamagnetic NPs are too small to drive the self-assembly, at least in weak magnetic fields (see Figure S3 in the SI for magnetic properties of the MNPs). However, the radial orientation of the mesogenic groups in the dendron shell favors the nematic director's radial anchoring. 
\rev{One of the crucial questions in describing the magnetic nanoparticles' dispersions is the coupling between the director in the bulk and the particles. In the case of micrometre-sized inclusions, the coupling occurs via anchoring of the director at the particles' surfaces and occurrence of the topological defects. The director's distortions caused by the particles' presence result in interactions tending to minimize the distortion energy. Those interactions drive the self-assembly of particle-defect states in the nematic matrices. When the particle size is on the nanometer scale, the anchoring is much weaker, and the defects-bound states are not expected. Still, the director's perturbations caused by the nanoparticles can drive the self-assembly by minimizing the distortion energy and maximizing the fluctuations.  
In case of our functionalized nanoparticles by the mesogenic-like dendron shell, the director is radially anchored at the MNPs surface (Figure~\ref{fig:fig4}a,b). To relieve the topological constraint, the dendron shell becomes deformed taking a tactoidal shape, which is similar to the case when a Saturn-ring defect loop collapses to a surface ring on the particle, in case of the weak anchoring~\cite{Stark.2001}. As demonstrated by theory,~\cite{Stark.2001, Lubensky:1998ei}  the surface ring configuration corresponds to the absolutely stable director configurations for small particles.  This deformation has a quadrupolar character and leads to the spontaneous self-assembly into kinked chains perpendicular to the mean director (Figure~\ref{fig:fig4}d)~\cite{Musevic:2006kw}. The director's bend regions overlap to reduce the elastic energy. Due to the smallness of the director deformation, we can use the far-field potential for a pair of interacting particles with quadrupole moments $c$, 
\begin{equation}
  U(\mathbf{r})=\frac{16 \pi K c^2}{9 r^5} (9 - 90 \cos^2\theta + 105 \cos^4\theta)
\end{equation}
where $r$ is the interparticle distance and $\theta$ is the angle with respect to $\mathbf{n}_0$ und $K$ is the effective elastic constant~\cite{Stark.2001}. The MNPs do not create defects, instead, the perturb the director field.
The director-mediated interactions can also lead to 2D or 3D structures,~\cite{Musevic:2006kw} but for short persistence lengths, 1D structures will be favourable. This results in a coupling between the director and the chain's orientation. }

Application of the magnetic field results in the magnetization of the superparamagnetic particles. Since the particles have isotropic shapes, there is no preferred direction for the chains. However, magnetic interactions of the particles give rise to the magnetic anisotropy. The chains magnetize easier when the field is parallel to the chain axis. As a result, the magnetic torque will align the chains along the field, and the director aligns obliquely (Figure~\ref{fig:fig4}e). Both torques have the same asymptotic field dependence proportional to $B^2$. Hence, an equilibrium tilt can be induced. The mean interparticle distance $d_{pp}$ is in the range of 95 nm (0.04 wt\%) and 55 nm (0.20 wt\%), which is in about ten molecular lengths of the 5CB mesogens. Thus, MNPs create mesoscopic confinement for the nematic. It is important to stress that it is the director mediated interactions and dendron-LC interactions (effective anchoring), that are responsible for  self-assembly. These interactions do not require a magnetic field. The magnetic chains are present already in the field-free state and improve the alignment of the director, which was confirmed by X-ray analysis.
%\subsection{Summary}

In summary, we demonstrated the manipulation of the optical properties of a soft colloidal LC-hybrid material using a delicate balance between the topologically-assisted colloidal self-assembly of MNPs and the anisotropic molecular interactions. The key feature in the design of these materials is the  functionalization of the MNPs, allowing an effective coupling between the magnetic particles and the LC matrix. 
The ferronematic  hybrid materials with dendron-functionalized MNPs showed a magnetically-induced tilt of the nematic director. We explain this tilt by the formation of  mesoscopic, topologically-driven self-assembled chain-like structures of MNPs stabilized through the long-range, director-mediated interactions. The magnetic alignment of these structures is responsible for the reorientation of the liquid crystal director.

In general, reorientation of mesogens and self-assembly of MNPs in such hybrid materials in response to electric or magnetic fields offer new avenues for tunable metamaterials or complex, hierarchical nanostructures which are unavailable by conventional synthesis. In particular, incorporating the MNPs in other types of LCs with chiral or smectic order could provide hybride materials exhibiting novel magneto-optical phenomena.

%In summary, we demonstrated the magnetically-induced tilt of the nematic director in a ferronematic with the dendron-functionalized MNPs. This tilt can be explained by the formation of the mesoscopic topologically driven self-assembled structures of MNPs stabilized through the long-range director-mediated interactions. 

%\begin{figure}[hbt]
%\centering
%	\includegraphics[width=0.8\columnwidth]{figs/structure.png}
%	\caption{Chemical structure (a) and the DSC diagrams (b) of the bent-core liquid crystal. the phase transitions are: Cr 102$^{\circ}$C (Col$_{\textrm{rec}}$ 83$^{\circ}$C) N$_{\textrm{CybC}}$ 143$^{\circ}$C Iso, where Col$_{\textrm{rec}}$ is a monotropic B$_{\textrm{1rev}}$-type phase, only observed on cooling or second heating: Cr, Col$_{\rm{rec}}$, N, and Iso represent crystal, columnar, nematic, and isotropic phases, respectively \cite{Alaasar:2013eo}. 
%		\label{fig:structure} 
%	}
%\end{figure}
%\subsection{Methods}
%methods
%\section{Results and Discussion}
%
%
%\section{Conclusions}
%conclusions

\subsection*{Acknowledgements} 
We thank the German Science Foundation (DFG) for financial support within the Priority Program (SPP1681) "Field controlled particle matrix interactions. Synthesis, multiscale modeling and application of magnetic hybrid materials",  within the  framework  of  Major  Research Instrumentation  Program  No.  INST  272/230-1, and project BE 2243/3. HN acknowledges the support of the German Science Foundation (DFG), Project NA 1668/1-1. The authors thank Prof. Ralf Stannarius for fruitful discussions and support.

\section*{Conflict of Interest}
The authors declare no conflict of interest.

\bibliography{WileyNJD-AMA}

\begin{thebibliography}{10}
\providecommand \doibase [0]{http://dx.doi.org/}%

\bibitem{Liu:2019co}
Liu X, Kent N, Ceballos A, et al.  {\it Science} 2019\string; 365(6450)\string:
  264--267.

\bibitem{Zhang:2019id}
Zhang X, Chen L, Lim KH, et al.  {\it Adv. Mater. Weinheim} 2019\string;
  31(11)\string: 1804540.

\bibitem{Mertelj:2014kv}
Mertelj A, Lisjak D, Drofenik M, Copic M.  {\it Nature} 2013\string;
  504(7479)\string: 237--241.

\bibitem{Wang:2014em}
Wang M, He L, Zorba S, Yin Y.  {\it Nano Lett} 2014\string; 14(7)\string:
  3966--3971.

\bibitem{Wang:2016iz}
Wang M, Yin Y.  {\it J. Am. Chem. Soc.} 2016\string; 138(20)\string:
  6315--6323.

\bibitem{Hu:2010ip}
Hu W, Zhao H, Shan L, et al.  {\it Liquid Crystals} 2010\string; 37(5)\string:
  563--569.

\bibitem{Rupnik:2015bi}
Rupnik PM, Lisjak D, Copic M, Mertelj A.  {\it Liquid Crystals} 2015\string;
  42(12)\string: 1684--1688.

\bibitem{deGennes:1992jj}
Gennes dPG.  {\it Angew. Chem. Int. Edit.} 1992\string; 104(7)\string:
  856--859.

\bibitem{Brochard:1970kp}
Brochard F, Gennes dPG.  {\it J. Phys. (France)} 1970\string; 31(7)\string:
  691--708.

\bibitem{Appel:2017hs}
Appel I, N{\'a}dasi H, Reitz C, et al.  {\it Phys. Chem. Chem. Phys.}
  2017\string; 504(19)\string: 1--9.

\bibitem{Mertelj:2018gq}
Potisk T, Mertelj A, Sebastian N, et al.  {\it Phys. Rev. E} 2018\string;
  97(1)\string: 254--18.

\bibitem{Koch:2020bt}
Koch K, Kundt M, Eremin A, Nadasi H, Schmidt AM.  {\it Phys. Chem. Chem. Phys.}
  2020\string; 22(4)\string: 2087--2097.

\bibitem{Draper:2011ej}
Draper M, Saez IM, Cowling SJ, et al.  {\it Adv. Funct. Mater.} 2011\string;
  21(7)\string: 1260--1278.

\bibitem{Nemati:2018cp}
Nemati A, Shadpour S, Querciagrossa L, et al.  {\it Nature Communications}
  2018\string; 9(1)\string: 1--13.

\bibitem{Seul_Science_1995}
Seul M, Andelman D.  {\it Science} 1995\string; 267(5197)\string: 476--483.

\bibitem{Chen_PRE_1995}
Chen CM, Lubensky TC, MacKintosh FC.  {\it Phys. Rev. E} 1995\string;
  51(1)\string: 504--513.

\bibitem{Kamien:2001uo}
Kamien RD, Selinger JV.  {\it COND.MAT.} 2001\string; 13(3)\string: R1--R22.

\bibitem{Hahsler:2020ci}
H{\"a}hsler M, Landers J, Nowack T, et al.  {\it Inorganic Chemistry}
  2020\string; 59(6)\string: 3677--3685.

\bibitem{Sun:2004jm}
Sun S, Zeng H, Robinson DB, et al.  {\it J. Am. Chem. Soc.} 2004\string;
  126(1)\string: 273--279.

\bibitem{Hahsler:2019kt}
H{\"a}hsler M, Behrens S.  {\it European Journal of Organic Chemistry}
  2019\string; 2019(48)\string: 7820--7830.

\bibitem{deGennes:1995vg}
Gennes dPG, Prost J. .
\newblock Clarendon Press .
\newblock 1995.

\bibitem{Kopcansky:2005cm}
Kop{\v c}ansk{\'{y}} P, Poto{\v c}ov{\'a} I, Konerack{\'a} M, et al.  {\it J
  Magn Magn Mater} 2005\string; 289\string: 101--104.

\bibitem{Tomasovicova:2008iq}
Toma{\v s}ovi{\v c}ov{\'a} N, Kop{\v c}ansk{\'{y}} P, Konerack{\'a} M, et al.
  {\it J Phys-Condens Mat} 2008\string; 20(20)\string: 204123--6.

\bibitem{Kopcansky:2008dm}
Kop{\v c}ansk{\'{y}} P, Toma{\v s}ovi{\v c}ov{\'a} N, Konerack{\'a} M, et al.
  {\it Phys. Rev. E} 2008\string; 78(1)\string: 011702--5.

\bibitem{Burylov:1995gc}
Burylov SV, Raikher YL.  {\it Molecular Crystals and Liquid Crystals Science
  and Technology. Section A. Molecular Crystals and Liquid Crystals}
  1995\string; 258(1)\string: 123--141.

\bibitem{Burylov:1995ie}
Burylov SV, Raikher YL.  {\it Molecular Crystals and Liquid Crystals Science
  and Technology. Section A. Molecular Crystals and Liquid Crystals}
  1995\string; 258(1)\string: 107--122.

\bibitem{Burylov:1990jz}
Burylov SV, Raikher YL.  {\it Physics Letters A} 1990\string; 149(5-6)\string:
  279--283.

\bibitem{Stark.2001}
Stark H.  {\it Physics Reports} 2001\string; 351(6)\string: 387--474.

\bibitem{Lubensky:1998ei}
Lubensky TC, Pettey D, Currier N, Stark H.  {\it Physical Review E}
  1998\string; 57(1)\string: 610 -- 625.

\bibitem{Musevic:2006kw}
Mu{\v s}evi{\v c} I, {\v{S}}karabot M, Tkalec U, Ravnik M, {\v Z}umer S.  {\it
  Science} 2006\string; 313(5789)\string: 954--958.

\end{thebibliography}
%\bibliography{papers.bib} 

%\nocite{*}% Show all bib entries - both cited and uncited; comment this line to view only cited bib entries;
%\bibliography{WileyNJD-AMA}%
%\begin{enumerate}[1]
%\item Use \verb"\bibliography{wileyNJD-AMA}" BST file for AMA reference style
%\item Use \verb"\bibliography{wileyNJD-APA}" BST file for APA reference style
%\item Use \verb"\bibliography{wileyNJD-AMS}" BST file for AMS reference style
%\item Use \verb"\bibliography{wileyNJD-VANCOUVER}" BST file for Vancouver reference style
%\item Use \verb"\bibliography{wileyNJD-ACS}" BST file for Chemistry reference style
%\end{enumerate}
%
%The normal commands for producing the reference list are:
%
%\begin{verbatim}
%\begin{thebibliography}{99}
%\bibitem{<x-ref label>}
%         <Reference details>
%.
%.
%.
%\end{thebibliography}
%\end{verbatim}

\end{document}